# A practical Laser-Heated Diamond Anvil Cell synthesis technique and recovery workflow for metastable $MnSb_2$ and $YbZn_2$ phases

S. Huyan [1,2†], R. F. S. Penacchio[1,3], D. Zhang[4], Z. Li[1,2], S. L. Morelhão[3], Raquel Ribeiro[1,2], P. C. Canfield[1,2], S. L. Bud'ko[1,2]

[1] *Ames National Laboratory, US DOE, Iowa State University, Ames, Iowa 50011, USA*
[2] *Department of Physics and Astronomy, Iowa State University, Ames, Iowa 50011, USA*
[3] *Institute of Physics, University of São Paulo, São Paulo, SP, Brazil*
[4] *Center for Advanced Radiation Sources, The University of Chicago, Chicago, Illinois 60637, USA*

[†] shuyan@iastate.edu

The creation and exploration of new materials under extreme pressure-temperature conditions has become increasingly reliant on laser-heated diamond anvil cell (LHDAC) techniques, which provide direct access to previously unexplored regions of multinary phase diagrams. Whereas numerous high-pressure phases have been identified in situ, systematic recovery and post-synthesis physical property characterization of these materials remain significant challenges. In this work, we present the development of an integrated LHDAC synthesis and demonstrate a practical LHDAC-based synthesis workflow that enables stabilization and recovery of metastable intermetallic phases for subsequent structural and transport studies. Using this approach, we successfully achieved LHDAC synthesis of high-pressure $MnSb_2$ and $YbZn_2$ phases under moderate pressures. Synchrotron X-ray diffraction and spatial mapping confirm dominant formation of the targeted phases, whereas laboratory-based refinement quantifies phase fractions despite intrinsic microstrain and minor secondary phases. High-pressure transport measurements on recovered samples reveal tunable by pressure electronic instabilities in both systems. In $MnSb_2$, pressure suppresses two high-temperature magnetic ordering anomalies, observed in transport, by 5 GPa and for higher pressures induces a new low-temperature feature that increases with further pressure increase. In hexagonal high-pressure $YbZn_2$, an electronic reconstruction emerges at ~11 GPa, characterized by semiconducting-like behavior from ~ 30 K to 300 K and a broad low-temperature coherence crossover near 30 K. Our results establish LHDAC synthesis not only as a structural discovery tool, but also as an experimental platform for investigating correlated quantum states stabilized far from equilibrium thermodynamic conditions.

## Introduction:

The exploration of materials under extreme pressure and temperature conditions has entered a transformative era with the development of laser-heated diamond anvil cell (LHDAC) techniques. [1-3]. These methods provide access to a broad static pressure-temperature phase space while also allowing for in-situ structural characterization using synchrotron X-ray diffraction, spectroscopy, as well as electrical transport and even thermodynamic measurements. [4,5] The earliest demonstrations of laser heating in DACs established the feasibility of reaching several thousand kelvin under multi-gigapascal pressures, [6,7] laying the foundation for high-pressure materials synthesis and mineral physics investigations.

Subsequent advances in optical design, double-sided heating geometries, beam shaping, and synchrotron integration have significantly improved temperature stability and spatial uniformity. [8,9] As a result, LHDAC has evolved into a powerful platform not only for probing deep-earth conditions, [3] but also for stabilizing novel condensed-matter phases inaccessible at ambient equilibrium. In particular, LHDAC-enabled synthesis has led to the discovery of record-breaking high-temperature superconducting superhydrides, such as $H_3S$ and $LaH_{10}$, with transition temperatures exceeding 200 K under megabar pressures. [10-12] Beyond superconductivity, laser-heating techniques in DACs have been instrumental in establishing phase stability, melting behavior, and thermophysical properties of deep-mantle and core materials at multi-megabar conditions. [13-15]

Despite these remarkable advances, a critical bottleneck remains: the reliable recovery, extraction, and comprehensive characterization of LHDAC-synthesized phases. The near-microscopic sample size, large thermal gradients during synthesis, accumulated microstrain, and mechanical constraints of the DAC environment often limit post-synthesis handling and ex-situ investigation. [16,17] As a result, many high-pressure discoveries rely primarily on in-situ structural signatures, while systematic physical property measurements on recovered samples remain relatively rare. Bridging this gap between extreme-condition synthesis and various physical properties characterization is essential for advancing high-pressure material research.

In this work, we report the development of an integrated LHDAC synthesis platform designed not only for high-temperature high-pressure phase synthesis but also for practical and efficient sample recovery workflow and subsequent characterization. To benchmark the approach, we selected two representative model systems: $MnSb_2$ and $YbZn_2$.

$MnSb_2$ crystallizes in the marcasite-type structure and can only be synthesized under high-pressure and high-temperature conditions. [18,19] Recent studies have demonstrated that, the high pressure phase $MnSb_2$ (hp-$MnSb_2$) develops multiple magnetic phase transitions (200 K and 118 K) upon cooling characterized by multi-sublattice, amplitude-modulated antiferromagnetic (AFM) order with temperature-dependent propagation vectors and inequivalent Mn sublattices. [19] First-principles calculations predict a sizable local moment on Mn, consistent with the experimentally observed ordered state. The coexistence of complex magnetic modulation and reduced symmetry makes $MnSb_2$ an appealing platform for examining the interplay between structural stabilization and electronic and magnetic ground states.

As for $YbZn_2$, it can be stabilized in a high-pressure hexagonal $MgZn_2$-type (*C*14) Laves phase (hp-$YbZn_2$) under high pressure. X-ray absorption spectroscopy and dynamical mean-field theory calculations indicate intermediate-valence behavior with a mean Yb valence of ~ 2.5, and a density of states at the Fermi level dominated by Yb 4*f* contributions. [20] These features make hexagonal $YbZn_2$ a representative correlated intermetallic system in which pressure stabilization may significantly influence electronic properties.

Notably, both $MnSb_2$ and $YbZn_2$ are metastable at ambient pressure, allowing us to demonstrate not only successful LHDAC synthesis but also structural verification after decompression and enabling independent transport measurements under pressure. By integrating high-pressure synthesis, quantitative spatial structural analysis, and subsequent electrical measurements, we establish a reproducible experimental route for studying metastable intermetallic phases. Rather than emphasizing system optimization, we focus on demonstrating feasibility and recoverability, thereby providing a practical pathway for linking LHDAC synthesis to magnetic and correlated-electron investigations.

## Method:

### Development of the laser-heating DAC system

An in-house laser-heated diamond anvil cell (LHDAC) system was developed for high-pressure and high-temperature synthesis. The system accommodates multiple diamond anvil cell geometries and employs a continuous-wave fiber laser (Coherent, model: Compact SE 100, with λ = 980 nm and maximum output power 100 W) as the heating source. The laser beam is focused to a spot size of approximately 30 μm to one side of the DAC using a custom optical assembly equipped with a CCD imaging system for real-time visualization. Temperature is monitored in situ by spectroradiometric measurements of the thermal emission collected from the laser heated area through the optical path of the laser-heating system. The emitted radiation is collected by a fiber-coupled spectrometer (Ocean Optics HR-6VN400) for temperature estimation based on blackbody radiation fitting. Under typical operating conditions, temperatures up to ~3000 K can be achieved, while static pressures approaching the megabar regime are accessible depending on the anvil configuration (i.e. culet sizes ≤ 200 um).

**Precursor preparation**

Polycrystalline precursor for $MnSb_2$ growth was prepared via solid-state reaction using elemental Mn and Sb in a molar ratio of 1:2. The mixture was sealed under vacuum in a silica tube in an $Al_2O_3$ crucible and heated to 785 °C at 100 °C/h for 24 h followed by air quenching of the still sealed tube. After cooling, the product was ground into powder for LHDAC use.

In a similar manner, $YbZn_2$ precursor was prepared from a Yb:Zn (1:2) mixture. The sealed ampoule was slowly heated to 850 °C over 10 h, held at 850 °C for 20 h, and then quenched in air. After cooling, the product was ground into powder for LHDAC use.

**High-pressure synthesis in LHDAC**

Two types of DACs were employed: a stainless-steel Mao-Bell-type cell [21] and a Be-Cu cell [22] compatible with a Quantum Design PPMS. Type Ia diamond anvils with 500 μm culets were used. Stainless-steel gaskets were pre-indented to ~80 μm thickness. A 500 μm hole was drilled and filled with a cubic BN-epoxy mixture, re-indented, and re-drilled to a final 200 μm diameter sample chamber. All pre-reacted precursors were ground into powder. A portion of the powders was pelletized, shaped into appropriate dimensions, and loaded into the sample chamber together with a few ruby spheres. The pressed NaCl (for $MnSb_2$ growth) and KCl (for $YbZn_2$ growth) thin layer pieces were inserted on both sides of the precursors and served as both pressure-transmitting media and thermal insulation. Pressure was determined by the ruby $R1$ fluorescence shift. [23,24] $MnSb_2$ was synthesized at 8.1 GPa and $YbZn_2$ at 9.6 GPa.

The DAC, loaded with the sample and pressure medium, was mounted on a three-axis (XYZ) micropositioning stage on the laser-heating optical line. Given that the laser has a focused beam spot size ~30 μm in diameter, to ensure a spatially uniform reaction across the sample chamber (~200 μm in diameter), a raster heating strategy was adopted. The sample area was divided into a 4 × 4 grid, and each position was sequentially aligned with the laser focus by translating the DAC using the micropositioning stage. At each position, the sample was heated for ~60 s with a laser power of ~32 W for $MnSb_2$ and ~55 W for $YbZn_2$. The heating process was monitored in real time via optical microscopy, where the laser-heated regions exhibited a distinct increase in brightness, indicative of elevated temperature. The raster scan over the full grid was repeated once to improve reaction homogeneity and promote more complete phase formation.  Currently the rastering is accomplished by hand and the complete, two pass, process takes roughly one hour.

Due to the localized nature of single-sided laser heating, significant axial and radial temperature gradients are expected within the sample chamber. As a result, the reaction is anticipated to be most complete in the directly heated regions, whereas areas farther from the laser focus may remain partially reacted or retain precursor-related phases.

**Synchrotron powder X-ray diffraction measurements**

Room-temperature high-pressure PXRD measurements were performed at beamline 13-BM-C (GSECARS), Advanced Photon Source (APS), Argonne National Laboratory (ANL). Monochromatic X-rays with a wavelength of 0.4283 Å were focused to a 10 μm × 10 μm spot. The LHDACs of as-grown $MnSb_2$ (SR006) and $YbZn_2$ (SR007) were directly mounted on the beamline setup for synchrotron PXRD measurements. Due to the large sample size, we adopted the cross-mapping scan strategy with a 10 um steps. More details will be discussed in the results part. The pressure of the sample was confirmed via the ruby fluorescence $R1$ line before the measurement. [23,24]

Two-dimensional diffraction images were integrated using the DIOPTAS, [25] and subsequent Rietveld refinements were carried out with GSAS-II. [26]

It is noteworthy that the Le Bail refinement was employed to extract lattice parameters, unit cell volume and to estimate relative phase fractions at each spatial position. The weighted profile *R*-factor ($R_{wp}$) provides a measure of the agreement between observed and calculated diffraction profiles; however, in the context of Le Bail analysis, it should not be interpreted as a unique validation of the structural model. This is particularly relevant for micro-focused high-pressure diffraction data, where limited powder averaging, preferred orientation, and overlapping reflections can allow multiple structural models to yield comparable $R_{wp}$ values. Therefore, structural assignment in this work is based not only on refinement quality but also on the consistency of peak positions, systematic absences, and agreement with known high-pressure phases reported in the literature.

**Sample recovery procedure**

After the synchrotron measurements, the DACs were decompressed to ambient pressure at room temperature for further study. In certain cases, decompression under low-temperature conditions can be employed to suppress kinetic relaxation processes and thereby improve retention of pressure-quenched metastable phases. Recovered samples were mechanically extracted from the gasket chamber. For $MnSb_2$, residual pressure medium (NaCl) was removed by rinsing in deionized water. For $YbZn_2$, due to its sensitivity to moisture, residual KCl was removed through repeated gentle rinsing in isopropanol (IPA) to minimize exposure to air and water. The recovered samples were then separated into small pieces for laboratory PXRD and transport measurements.

**In-house powder X-ray diffraction**

The recovered samples were characterized using a  equipped with a silver radiation source ($K_\alpha$ avg. ~ 0.56087 Å) in transmission mode, operating at 65 kV and 0.67 mA. The sample extracted from the DAC was shaped into ~ 50 um size and mounted on the Nylon loop single crystal mount with N grease and subsequently placed on the goniometer for measurement. The beam spot size of the equipment is ~ 100 µm. Compared to synchrotron measurements, broader diffraction peaks were observed, attributed to a) Shorter sample-to-detector distance (~32 mm) and the microstrain and non-hydrostatic stress accumulated during decompression. The data were collected as two-dimensional CCD image and powder reduction was performed using CrysAlisPro (Rigaku OD, 2023), and subsequent Rietveld refinements were carried out with GSAS-II. [26] Microstrain parameters were incorporated during refinement to improve fitting reliability and enable quantitative phase fraction determination.

**Electrical transport measurements under high pressure**

Post synthesis and post extraction from the reaction LHDAC, the electrical resistance was measured up to 15 GPa using a Be-Cu DAC [22] compatible with a Quantum Design Physical Property Measurement System (PPMS). The cell employed Type-la diamond anvils with 500 μm culet size and a stainless-steel gasket. After pre-indenting the gasket to ~10 GPa, a concentric 500 μm hole was drilled by electric-discharge machining (EDM) and back-filled with a cured mixture of cubic BN powder and epoxy. This insert was then compressed to 10 GPa again and re-drilled using mechanical micro-drilling machine to a 200 μm diameter hole to form the sample chamber.

The bottom space of the sample chamber was first covered with a thin NaCl layer along with a small ruby sphere. $MnSb_2$ and $YbZn_2$ samples extracted after synthesis were shaped into rectangles and placed above this layer, and the Pt foil electrodes, were arranged in a van der Pauw configuration on the pressed sample. An additional thin NaCl layer was placed over the sample and electrodes to improve the hydrostaticity at high pressure. Pressure was determined from the ruby *R*1 fluorescence line [23,24].

## Results and discussion:

After the synthesis was done, the LHDAC reaction-cells were carried to APS Sector 13-BM-C, and the high-pressure synchrotron PXRD measurements were performed to verify phase formation in-situ, directly after laser heating growth and without decompression. This is critical because many pressure-stabilized intermetallics can partially or completely back-transform upon unloading. Therefore, structural verification under synthesis conditions provides a reliable benchmark for evaluating reaction completeness and guiding subsequent recovery.

To assess the spatial homogeneity of the reaction, cross-shaped micro-focused diffraction mapping (10 μm step size) was performed across the sample chamber, as illustrated in Figs. 1 & 2 for $MnSb_2$ (SR006) and $YbZn_2$ (SR007), respectively, where the relative targeted phase fraction at each point is shown by the color contrast (Figs. 1a & 2a). Representative diffraction patterns collected along orthogonal directions are displayed in Figs. 1d,e & 2d,e; the spatial evolution of the unit-cell volume, and weighted profile *R*-factor ($R_{wp}$) are summarized in Figs. 1b,c & 2b,c. These spatially resolved measurements reveal that the degree of reaction varies noticeably on a length scale of ~10 μm, reflecting the combined effects of axial thermal gradients inherent to single-sided laser heating and lateral inhomogeneity arising from the raster-based heating approach.

Despite this spatial inhomogeneity, a key result concluding from the mapping is that a substantial fraction of the sample volume corresponds to the targeted phase. Averaging over the scanned positions indicates that approximately 40% or more of the probed regions are dominated by the high-pressure phase for both systems. Importantly, although the raster-based heating strategy is designed to approximate spatially uniform heating, the spatial maps reveal inhomogeneity in the phase distribution. This indicates that the local reaction conditions might be governed by a complex interplay between laser absorption, thermal conduction, and stress heterogeneity within the DAC environment.

Given the spatial variability revealed by the mapping, it is instructive to examine diffraction patterns from regions exhibiting the best phase purity. Fig. 3 presents representative diffraction data collected from such "best-case" regions, corresponding to visually darker areas in the optical images in Figs 1a & 2a that indicate stronger laser heating and more complete reaction.

For $MnSb_2$ (SR006), the selected diffraction pattern (Fig. 3a) is dominated by reflections that can be indexed to the high-pressure marcasite-type structure (*Pnnm*, SG #58), although diffraction peaks from likely unreacted Sb and MnSb remain detectable. This confirms that locally high phase purity can be achieved within the LHDAC-synthesized sample, even though the overall reaction remains spatially inhomogeneous. In addition, the relative phase fraction indicates that the targeted phase is significantly lower on the left side of the mapping region. This region corresponds to the area of outside sample's boundary where the signal is dominated in part by the NaCl medium.

Similarly, for $YbZn_2$ (SR007), the diffraction pattern shown in Fig. 3b corresponds to a region with the highest observed phase purity. The majority of reflections are consistent with the high-pressure hexagonal $MgZn_2$-type Laves phase ($P6_3/mmc$, SG #194), with contributions from low-pressure (lp-) $YbZn_2$ and $Yb_3Zn_{11}$ second phases. A quantitative estimate based on Le Bail refinement yields approximately 62.9 mol% hp-$YbZn_2$, 20.6 mol% lp-$YbZn_2$, and 16.6 mol% $Yb_3Zn_{11}$.

All above results demonstrate that, although the reaction is not spatially uniform, the LHDAC synthesis produces the desired phase over a substantial portion of the sample chamber, and that spatially resolved diffraction mapping provides a useful means to assess phase homogeneity and guide targeted secondary heating to improve phase purity.

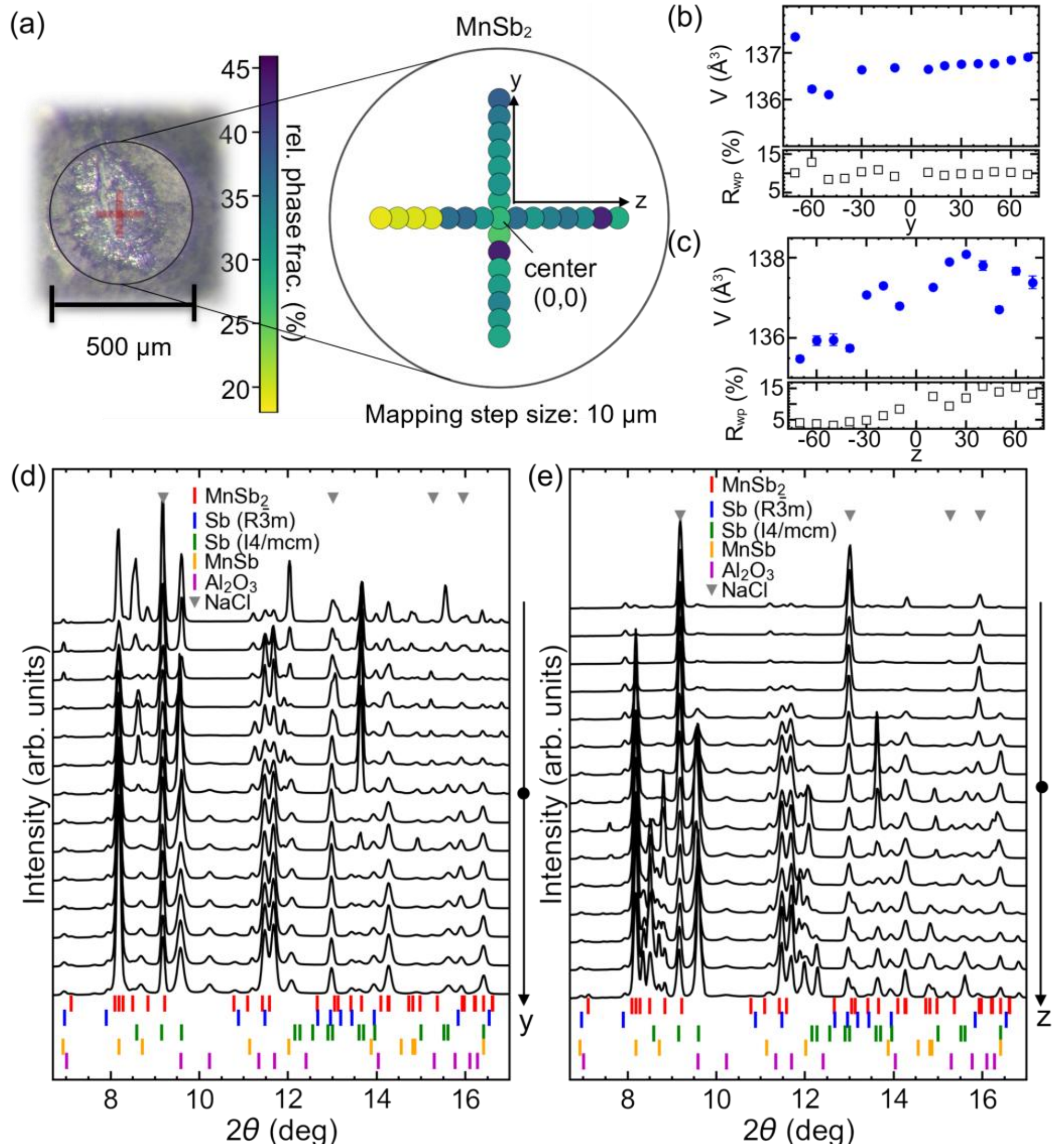

Fig. 1 Synchrotron powder X-ray diffraction spatial mapping using a cross-scanning strategy. (a) Schematic illustration of cross-shaped scanning geometry. The left panel shows a real-time optical image of the DAC sample chamber. The red cross stands for the mapping region. The color scale demonstrates the relative fraction of the main phase (hp-$MnSb_2$) at selected positions indicated within the central circular region (right panel). The relative phase fractions were roughly estimated from the product of sum of LeBail intensities and refined unit cell volumes. Here, we need to roughly account for the very different scatter densities between the $MnSb_2$, MnSb, low-pressure and high-pressure Sb phases. Note that the phase fractions reported here are intended to reflect the relative spatial distribution of the high-pressure phase, whereas the absolute value is not very accurate. The refined unit-cell volume of hp-$MnSb_2$ and its weighted profile *R*-factor ($R_{wp}$) are plotted as a function of position along the (b) *y*- and (c) *z*- directions. The corresponding diffraction patterns collected along the y- and z-direction are shown in (d) and (e), respectively. The incident X-ray wavelength was 0.4283 Å.

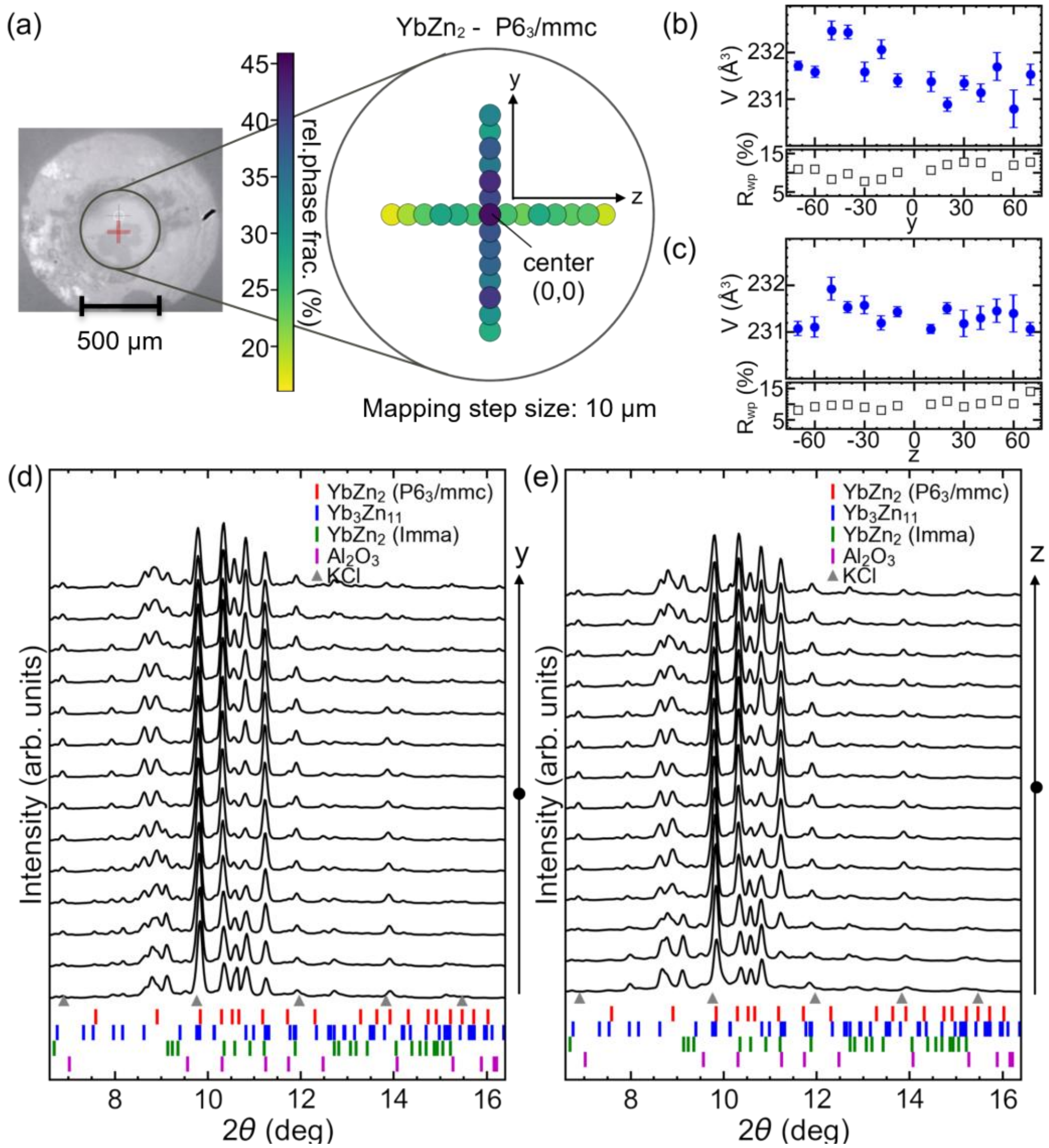

Fig. 2 Synchrotron powder X-ray diffraction spatial mapping using a cross-scanning strategy. (a) Schematic illustration of cross-shaped scanning geometry. The left panel shows a real-time optical image of the DAC sample chamber. The red cross stands for the mapping region. The color scale demonstrates the relative fraction of the main phase (hp-$YbZn_2$) at selected positions indicated within the central circular region (right panel). The relative phase fractions (rel. phase frac.) were roughly estimated by normalizing the sum of LeBail intensities. Here, lp-$YbZn_2$, hp-$YbZn_2$ and $Yb_3Zn_{11}$ have roughly same scatter density, and don't need correction by unit cell volume. Note that the phase fractions reported here are intended to reflect the relative spatial distribution of the high-pressure phase, whereas the absolute value is not very accurate. The refined unit-cell volume of hp-$YbZn_2$ and its weighted profile *R*-factor ($R_{wp}$) are plotted as a function of position along the (b) *y*- and (c) *z*- directions. The corresponding diffraction patterns collected along the y- and z-direction are shown in (d) and (e), respectively. The incident X-ray wavelength was 0.4283 Å.

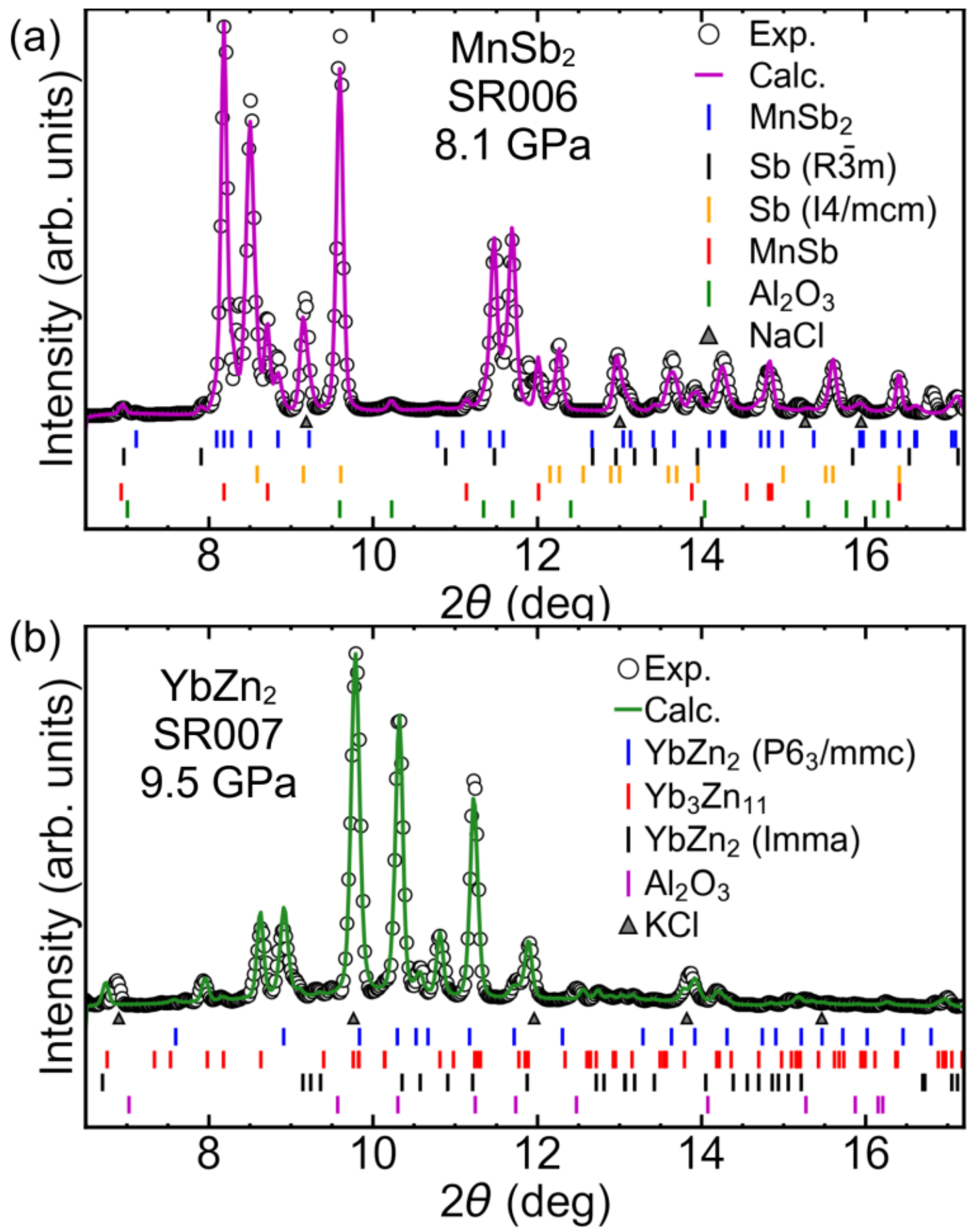

Fig. 3 Synchrotron PXRD patterns collected from the best phase-purity locations identified in the spatial mapping for post-growth (a) $MnSb_2$ (SR006, 8.1 GPa) and (b) $YbZn_2$ (SR007, 9.5 GPa). The incident X-ray wavelength was 0.4283 Å.

It is also important to emphasize that these "best-case" patterns shown in Fig. 3 represent locally optimized regions rather than the average behavior across the sample chamber. Taken together with the spatial mapping results, this indicates that the LHDAC synthesis produces a spatially inhomogeneous sample, in which well-reacted regions coexist with less-reacted areas.

For subsequent transport measurements, knowledge of the phase purity and distribution is essential due to the spatial inhomogeneity. Therefore, identification of the spatial region where the target phase predominates is necessary for reliable downstream characterization. Following the synchrotron measurements, samples were recovered from the DACs, manually picked, shaped into small pieces (~ 50 μm), and characterized with PXRD mode using the in-house Rigaku Synergy diffractometer equipped with a silver radiation source, *Kα* avg. ~ 0.560 Å. Compared to the high-resolution synchrotron data collected at APS 13-BMC, the laboratory diffraction patterns exhibit noticeably broader peaks, as shown in Fig. 4 and Fig. 5. Such peak broadening can arise from the small sample to detector distance of the in-house system (~32 mm) compared with much larger distance in synchrotron (~194 mm) and the non-hydrostatic stress and lattice distortion accumulated during decompression which may further contribute to strain-induced broadening.

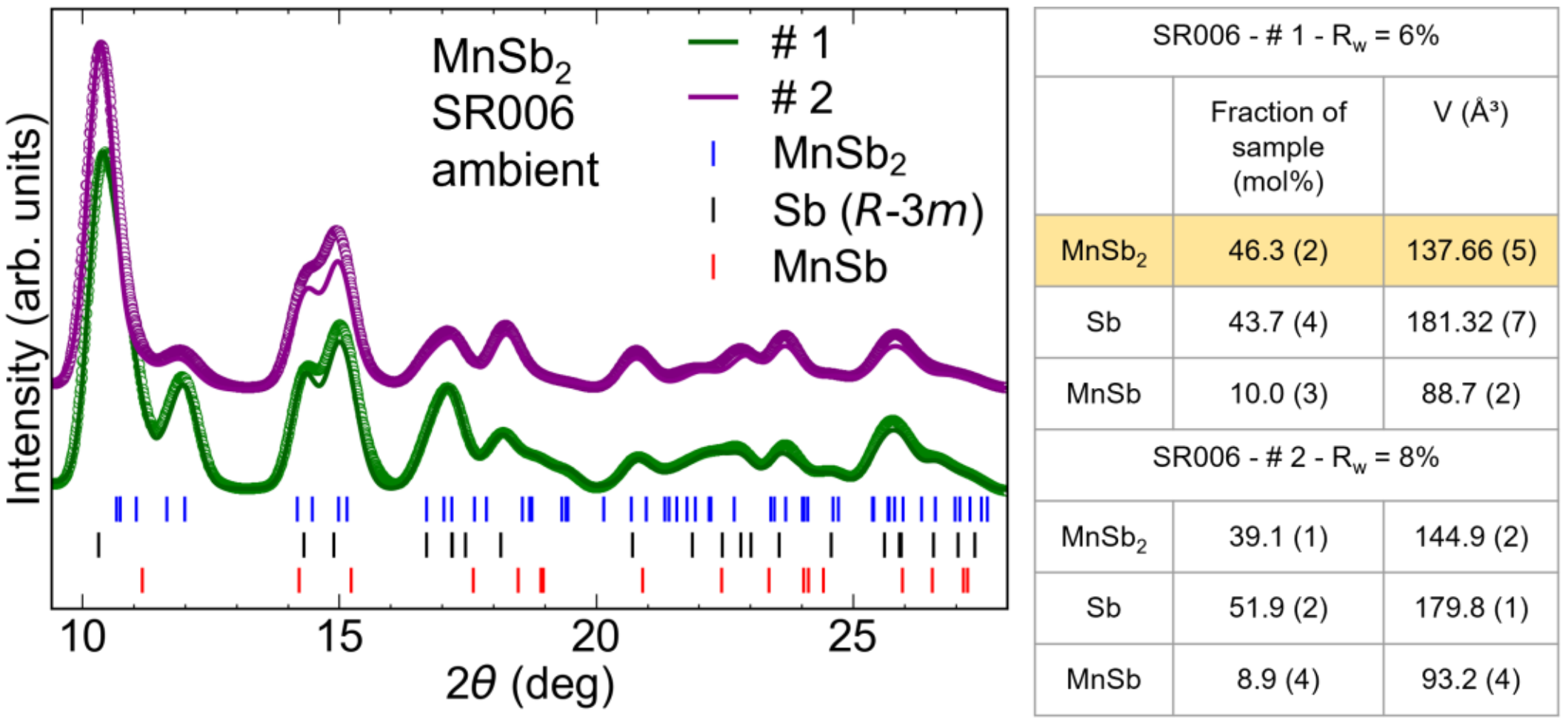

| SR006 - # 1 - $R_w$ = 6% | | |
|---|---|---|
| | Fraction of sample (mol%) | V (Å³) |
| $MnSb_2$ | 46.3 (2) | 137.66 (5) |
| Sb | 43.7 (4) | 181.32 (7) |
| MnSb | 10.0 (3) | 88.7 (2) |
| SR006 - # 2 - $R_w$ = 8% | | |
| $MnSb_2$ | 39.1 (1) | 144.9 (2) |
| Sb | 51.9 (2) | 179.8 (1) |
| MnSb | 8.9 (4) | 93.2 (4) |

Fig. 4 Powder X-ray diffraction patterns from two selected recovered $MnSb_2$ pieces measured after decompression using a Rigaku Synergy diffractometer (integrated from two-dimensional diffraction images). Silver radiation source with *Kα* avg. ~ 0.560 Å is used. Additional reflections arise from Sb and MnSb phases. The table on the right summarizes the refined phase fractions and unit-cell volumes of each phase. Sample #1 was selected for subsequent electrical transport measurements.

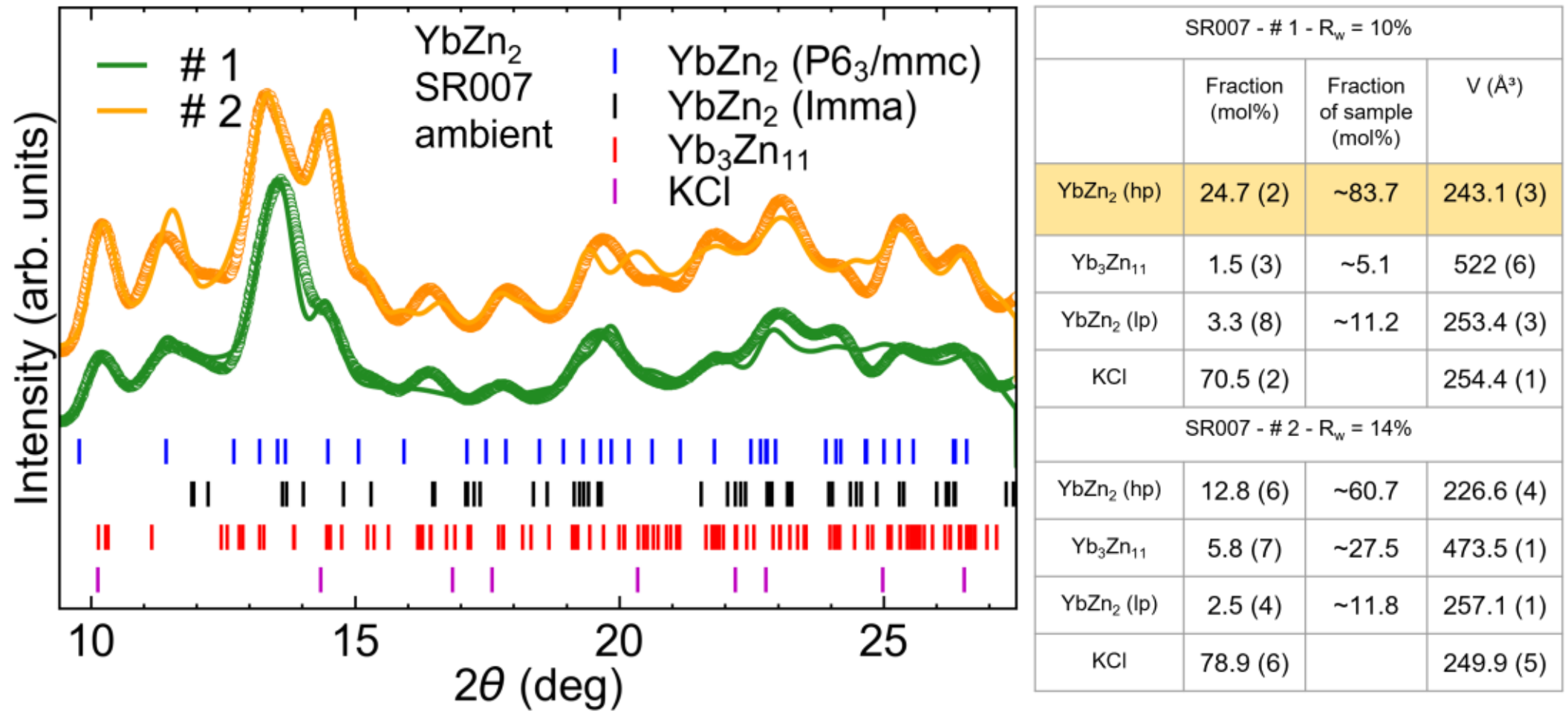

| SR007 - # 1 - $R_w$ = 10% | | | |
|---|---|---|---|
| | Fraction (mol%) | Fraction of sample (mol%) | V (Å³) |
| $YbZn_2$ (hp) | 24.7 (2) | ~83.7 | 243.1 (3) |
| $Yb_3Zn_{11}$ | 1.5 (3) | ~5.1 | 522 (6) |
| $YbZn_2$ (lp) | 3.3 (8) | ~11.2 | 253.4 (3) |
| KCl | 70.5 (2) | | 254.4 (1) |
| SR007 - # 2 - $R_w$ = 14% | | | |
| $YbZn_2$ (hp) | 12.8 (6) | ~60.7 | 226.6 (4) |
| $Yb_3Zn_{11}$ | 5.8 (7) | ~27.5 | 473.5 (1) |
| $YbZn_2$ (lp) | 2.5 (4) | ~11.8 | 257.1 (1) |
| KCl | 78.9 (6) | | 249.9 (5) |

Fig. 5 Powder X-ray diffraction patterns from two selected recovered $YbZn_2$ samples measured after decompression using a Rigaku Synergy diffractometer with silver radiation source with $K\alpha$ avg. ~ 0.560 Å. The patterns were obtained by integrating two-dimensional diffraction images. Additional reflections arise from the pressure-transmitting medium KCl, as well as lp-$YbZn_2$ and $Yb_3Zn_{11}$ phases. The table on the right summarizes the refined phase fractions and unit-cell volumes. Sample #1 was selected for subsequent electrical transport measurements, whereas Sample #2 exhibits less reliable lattice-parameter refinement due to overlapping reflections and other-phases contributions.

Despite the broader diffraction features, the main structural phase can still be unambiguously identified, as shown in Fig. 4 and Fig. 5. The diffraction pattern contains reflections consistent with the targeted high-pressure phase, together with additional peaks arising from secondary or unreacted phases, indicating incomplete phase purity. During Rietveld refinement, microstrain effects were explicitly incorporated into the structural model, which significantly improved the fitting quality and allowed reliable extraction of lattice parameters and quantitative phase fractions. The refined phase percentages are summarized in the tables in Fig. 4 and Fig. 5. This analysis enabled us to select samples with relatively higher phase purity for subsequent transport measurements.

This step is critical within the LHDAC workflow: by quantitatively evaluating phase composition after recovery, samples with higher phase purity can be selected for subsequent transport measurements, ensuring that electronic properties predominantly reflect the targeted metastable phase. The $MnSb_2$ sample has a large amount of elemental Sb present, but Sb will not contribute signatures of potential magnetic ordering or related phase transitions. It can become superconducting at low temperatures for high pressures, as is the case shown below.

We performed high-pressure electrical resistance measurements on polycrystalline $MnSb_2$ recovered from LHDAC synthesis and reloaded into a DAC for measurement up to ~15 GPa. At ambient pressure, the temperature-dependent resistance shows metallic behavior with a convex shape, which qualitatively resembles the results of recent measurements on $MnSb_2$ single crystal at ambient pressure. [19] Although the residual resistance ratio (RRR $\approx$ 6) is significantly lower compared to the single crystals (RRR $\approx$ 60), the overall temperature dependence remains consistent, indicating that the electronic ground state is preserved despite enhanced scattering from grain boundaries and pressure-induced lattice strain. Remarkably, subtle resistive anomalies are discernible in the d$R$/d$T$ plot, as shown in Fig. 6b, and are consistent with the two magnetic phase transitions ($T_1$ and $T_2$) that were found for $MnSb_2$ at ambient pressure. [19] There is also a broad cross over feature in d$R$/d$T$ that we label $T$*.

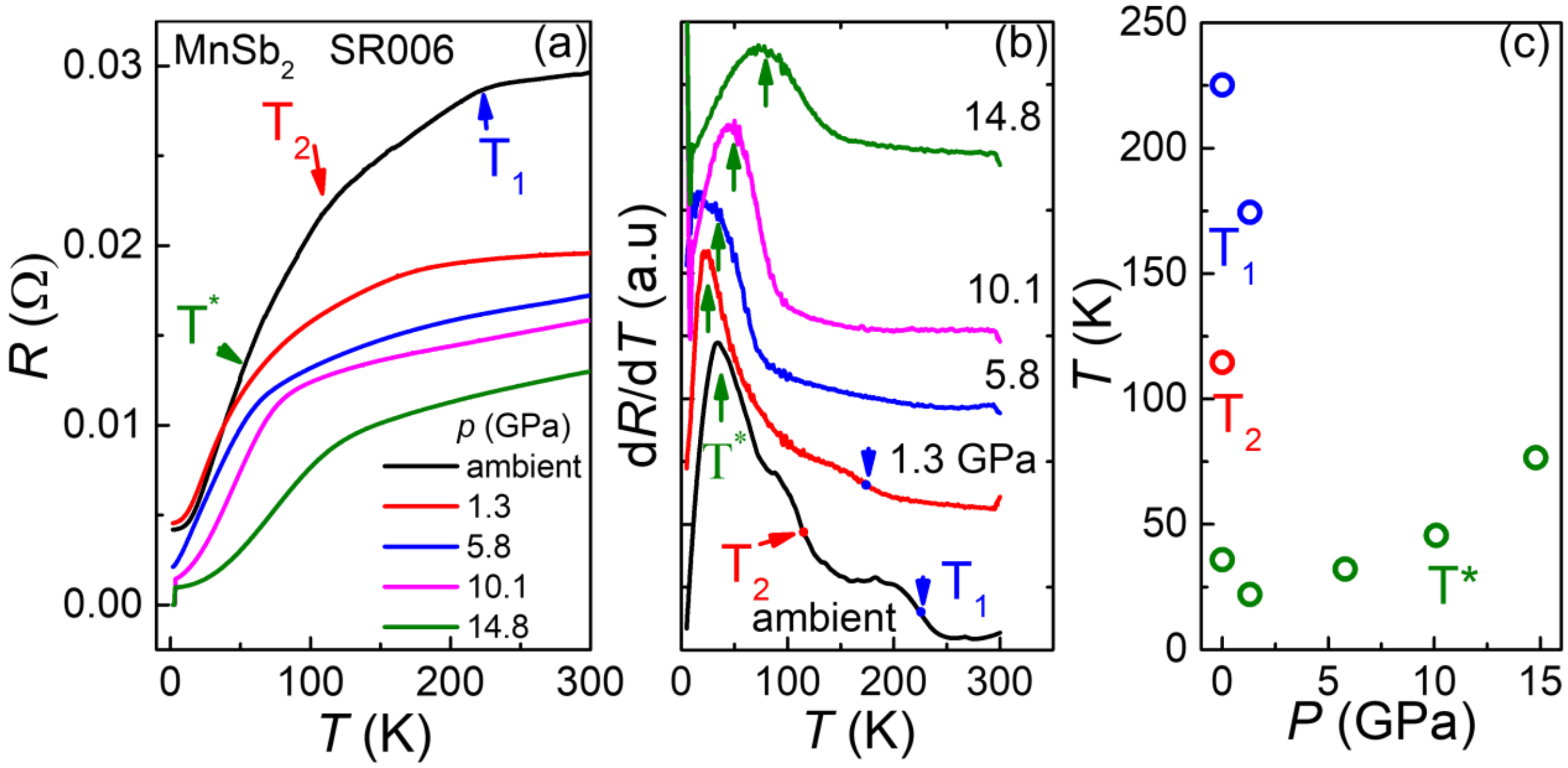

Fig. 6 (a) Temperature dependence of the resistance $R(T)$ for $MnSb_2$ (SR006, sample #1) at various pressures. The blue, red, and green arrows indicate the characteristic anomalies denoted as $T_1$, $T_2$, and $T$*, respectively. (b) Temperature derivative d$R$/d$T$ as a function of temperature. The criteria for determining $T_1$, $T_2$, and $T$*, are indicated in panel (b), defined by the midpoint of the shoulder features and the position of the corresponding peak. (c) Pressure dependence of $T_1$, $T_2$, and $T$*.

Upon increasing pressure, $T_1$ and $T_2$ are rapidly suppressed and become indiscernible above ~1.3 GPa. This rapid disappearance implies that the associated instability is highly sensitive to lattice compression and likely tied to fine details of the electronic structure. As a result, the high-temperature linear resistance region extends to lower and lower temperatures as pressure is initially increased. Interestingly, a pronounced low-temperature downturn in resistance emerges under pressure and becomes increasingly prominent with further compression. The characteristic temperature $T$*, extracted from the inflection in $R(T)$ or corresponding feature in d$R$/d$T$, increases monotonically with pressure from 1.3 GPa to 14.8 GPa. The systematic enhancement of $T$* suggests that pressure stabilizes a new electronic/magnetic state at low temperature. Importantly, the character of this feature is manifested as a change in slope rather than a discontinuity, indicating that the underlying transformation is either a broadened (magnetic) phase transition or some temperature driven cross-over. The non-hydrostatic pressure condition due to the use of solid pressure transmitting medium could be part of the reason for the lack of sharp features of the transition-like behavior.

In addition to the pressure-induced anomalies discussed above, a clear superconducting-like transition is observed in the resistance curves and above 10.1 GPa. The resistance drops sharply at low temperature and reaches zero with superconducting $T_c$ ~ 3.6 K at 10.1 GPa. $T_c$ decreases a little to ~3.4 K at around 14.8 GPa. Careful comparison with the known pressure-temperature phase diagram of elemental Sb reveals that the observed superconducting transition closely matches that of Sb under pressure, [27] consistent with our structural analysis indicating the coexistence of residual Sb with the $MnSb_2$ phase.

Recent density functional theory calculations predict that marcasite-type $MnSb_2$ favors an AFM ground state with a sizable local moment (~2.8 μB per Mn), consistent with experimental observations of long-range magnetic order in the single crystal sample [19]. Neutron diffraction measurements reveal a complex multi-sublattice, amplitude-modulated AFM structure with temperature-dependent propagation vectors and inequivalent Mn sublattices. The coexistence of sizable local moments and metallic transport suggests that $MnSb_2$ lies in a delicate regime where magnetic exchange interactions, electronic bandwidth, and Mn-Sb hybridization are strongly intertwined. In such scenario, moderate lattice compression can significantly alter the magnetic exchange interactions and electronic bandwidth, thereby tuning the balance among competing magnetic configurations within the ordered state. The pressure-induced suppression of $T_1$ and $T_2$, together with the emergence and monotonic enhancement of $T^*$, may therefore reflect a pressure-driven evolution of this complex magnetic ground state.

While the present transport data alone cannot unambiguously distinguish between magnetic ordering and a purely Fermi surface topology reconstruction in $MnSb_2$ system, the systematic pressure evolution and the monotonic increase of $T^*$ strongly support the presence of a pressure-stabilized collective ground state. Further probes like high-pressure magnetic susceptibility, neutron diffraction, or Hall measurements on a single crystalline sample will be essential to determine the microscopic origin of this transition.

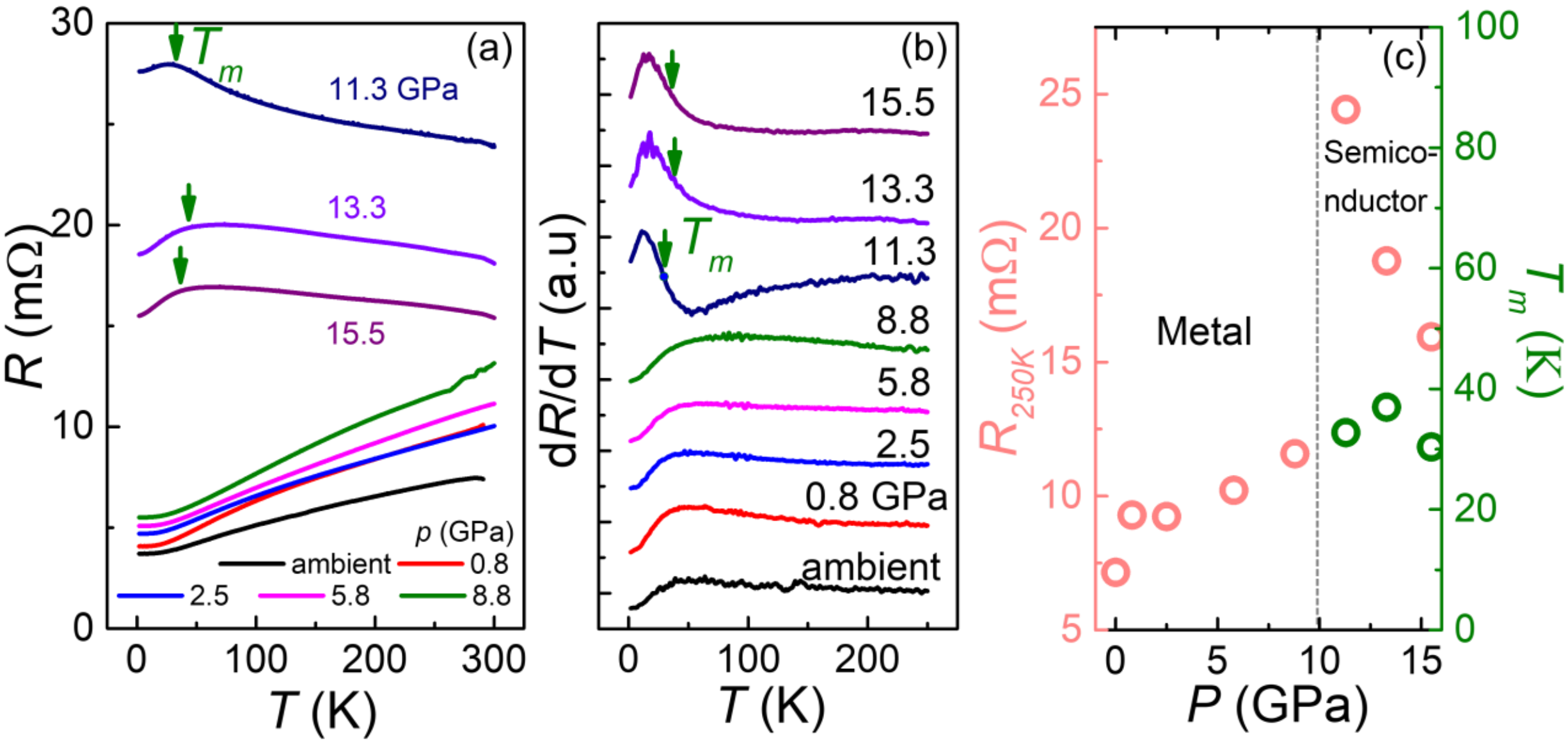

Fig. 7 (a) Temperature dependence of the resistance $R(T)$ for $YbZn_2$ (SR007, sample #1) at various pressures. The green arrows indicate the characteristic anomaly denoted as $T_m$. (b) Temperature derivative d$R$/d$T$ as a function of temperature. The criterion for determining $T_m$ is indicated in panel (b), defined as the midpoint of the shoulder-like feature. (c) Pressure dependence of the resistance at 250 K (left axis) and $T_m$ (right axis).

Electrical transport measurements were also carried out on the recovered hp-$YbZn_2$, which was subsequently reloaded into a diamond anvil cell for compression up to 15 GPa. As reported previously, the hexagonal phase shows metallic behavior over the entire 2-300 K range without any visible phase transitions and follows a Fermi-liquid-like dependence, $\rho = \rho_0 + AT^2$ at low temperatures. [20] In our measurement, as shown in Fig. 7a and Fig. 8, the metallic, Fermi-liquid behavior (in our case, $R = R_0 + AT^2$) persists up to 8.8 GPa. The residual resistance $R_0$increases monotonically with pressure, while the coefficient $A$ remains nearly pressure independent. This evolution indicates that moderate lattice contraction enhances elastic scattering but does not immediately destabilize the coherent quasiparticle state. Although RRR at 1.8 K is ~ 1.7 at ambient condition, which is smaller than the reported value (~3), reflecting enhanced residual scattering in our polycrystalline specimen, the overall temperature dependence resistance is qualitatively identical.

Upon further increasing the pressure, a pronounced change emerges near 11.3 GPa (Fig. 7a). At this pressure, the overall resistance exhibits an abrupt increase in size (increasing by roughly 2-5 times, depending on temperature), and the slope of resistance curve changes from positive to negative over a broad temperature interval, giving rise to semiconducting-like behavior at intermediate temperatures, as shown in Figs. 7a,b. This evolution contrasts sharply with the robust metallic state observed at low pressure range and strongly suggests an intrinsic pressure-driven modification of the electronic structure rather than a simple enhancement of scattering.

Notably, despite the semiconducting-like behavior at intermediate temperatures, a broad low-temperature downturn develops below ~30 K, labeled as $T_m$ as shown in Fig. 7a,b where the resistance decreases upon further cooling. This crossover-like feature persists upon further compression to 15 GPa. While the overall resistance magnitude is gradually reduced at higher pressures, the characteristic temperature scale and the broad nature of the low-temperature downturn remain nearly unchanged. Such behavior suggests a pressure-induced two-regime transport scenario. Above ~30 K, the semiconducting-like temperature dependence indicates a strongly scattered electronic state, possibly associated with partial gap formation or incoherent 4*f*-conduction electron hybridization. In contrast, the low-temperature resistance downturn reflects the recovery of metallic-like transport or a loss of some, possibly magnetic, scattering below a well-defined temperature scale.

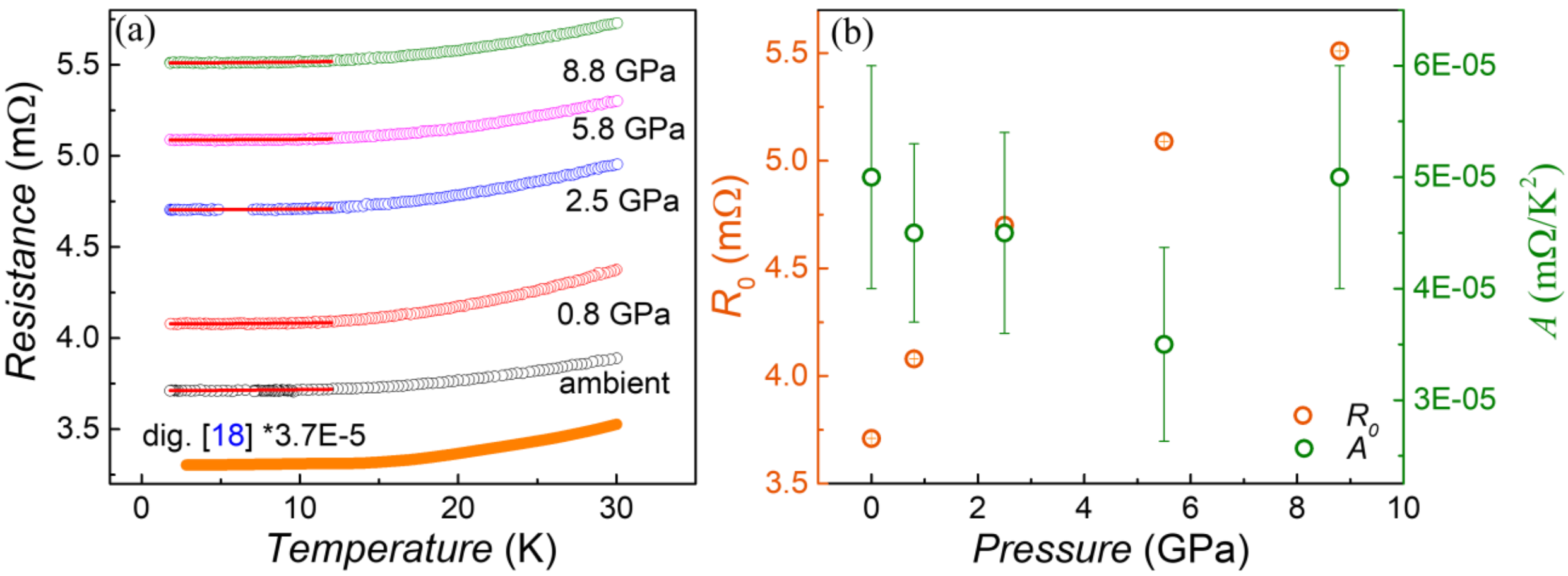

Fig.8 (a) Power-law fits to the low-temperature $R(T)$ curves of $YbZb_2$ at various pressures. The data in the temperature range 1.8 - 12 K were fitted using the Fermi-liquid form,: $R(T) = R_0 + AT^2$. The red curves represent the fitted results. The orange solid curve is digitized from Ref. [20], its resistivity values were scaled by a factor of 3.7E-5 for clarity and direct comparison with the present measurements. (b) Pressure dependence of the fitted Fermi-liquid parameters.

Two interpretations may account for this evolution. One scenario involves the gradual development of a correlated quasiparticle state under pressure. In this picture, the intermediate-temperature semiconducting-like response corresponds to an incoherent regime, while the low-temperature downturn marks the onset of coherent heavy-electron transport driven by enhanced *f*-conduction electron hybridization.

Alternatively, if the apparent broadness of the resistive anomaly near ~30 K arises from non-hydrostatic conditions associated with the use of a solid PTM as well as possibly inhomogeneous / locally stressed sample, the underlying transition may in fact be sharper and could correspond to the onset of magnetic ordering. Yb-based systems are often located near valence and magnetic instabilities, and pressure could stabilize a magnetically ordered phase. Such ordering would naturally lead to partial Fermi-surface reconstruction or superzone-gap formation, giving rise to semiconducting-like behavior above the ordering temperature.

The relatively broad nature of the feature and the absence of clear discontinuity prevent a definitive distinction between a continuous crossover and a second-order phase transition. Therefore, whereas the transport data strongly indicate a pressure-induced electronic reconstruction in hp-$YbZn_2$, complementary magnetic and thermodynamic measurements under improved hydrostatic conditions are required to clarify the microscopic origin of this anomaly. Taken together, these results demonstrate that the correlated electronic state of hp-$YbZn_2$ is highly tunable by pressure.

## Conclusion:

In conclusion, we have developed and implemented an integrated LHDAC synthesis and recovery platform capable of stabilizing metastable high-pressure intermetallic phases and enabling their subsequent physical properties characterization. By combining controlled laser heating, spatially resolved synchrotron diffraction mapping, quantitative laboratory diffraction, and pressure-dependent transport measurements, we establish a reproducible experimental workflow that bridges extreme-condition synthesis by LHDAC with correlated-electron investigations.

Using this workflow, we successfully synthesized and recovered metastable hp-$MnSb_2$ and hp-$YbZn_2$ phases. Synchrotron measurements confirm dominant formation of the targeted high-pressure phases. Crucially, laboratory-based PXRD characterization after recovery allows quantitative evaluation of phase fractions despite peak broadening arising from microstrain and non-hydrostatic stress. This step serves as a key bridge between in-situ synthesis and ex-situ physical property measurements, enabling informed sample selection and ensuring that the observed electronic behavior predominantly reflects the intended metastable phase.

Transport measurements reveal that both systems host electronically delicate states that are highly sensitive to lattice compression. In $MnSb_2$, pressure rapidly suppresses ambient-pressure magnetic ordering temperatures and stabilizes a distinct low-temperature correlated state, consistent with the proximity of competing magnetic and itinerant configurations predicted by theory. The monotonic enhancement of $T^*$ under pressure indicates that modest lattice contraction is sufficient to shift the balance between nearly degenerate electronic ground states in marcasite family. In hp-$YbZn_2$, a pronounced electronic reconstruction occurs near ~11 GPa, marked by an abrupt increase in resistance and a reversal of the temperature coefficient of resistance (metallic to semiconducting behavior). The emergence of semiconducting-like behavior and together with low-temperature metallic recovery, signals the formation of competing correlated regimes. Whether this anomaly reflects the onset of coherent heavy-electron behavior or pressure-stabilized magnetic ordering, the data demonstrates that hp-$YbZn_2$ resides near a critical electronic instability that can be accessed through moderate compression.

More broadly, this work demonstrates that LHDAC synthesis can evolve from a structural discovery technique into a practical platform for engineering and probing metastable quantum materials. By integrating spatial structural verification, quantitative post-recovery phase analysis, and downstream electrical characterization, our approach establishes a practical route for recovering LHDAC-synthesized materials and enabling comprehensive ex-situ investigations beyond in-situ structural identification.

## Acknowledgement

Development of capabilities for synthesis and measurements under pressure was supported by Ames National Laboratory's Laboratory Directed Research and Development (LDRD) program. Work at Ames National Laboratory is supported by the US DOE, Basic Energy Sciences, Material Science and Engineering Division under contract no. DE-AC02-07CH11358. RFSP’s one-year visit to Ames Laboratory and Iowa State University was supported by the São Paulo Research Foundation (FAPESP), Brazil, Process Number 2024/08497-6. RFSP also acknowledges support from FAPESP under Process Number 2021/01004-6. SLM acknowledges support from FAPESP under Process Number 2023/10775-1. Synchrotron experiments were performed at GeoSoilEnviroCARS (The University of Chicago, Sector 13), Advanced Photon Source, Argonne National Laboratory. GeoSoilEnviroCARS is supported by the National Science Foundation–Earth Sciences via SEES: Synchrotron Earth and Environmental Science (EAR–2223273).

## Competing interests

The authors declare no competing interests.

## Data Availability

The data that support the findings of this study will be made available in DataShare, an open-access repository at Iowa State University upon acceptance of the manuscript.